\documentclass[11pt]{article}
\usepackage[a4paper]{geometry}
\usepackage{amsmath}%
\usepackage{amsfonts}%
\usepackage{amssymb}%
\usepackage{graphicx}
\usepackage{float}
\numberwithin{equation}{section}
\restylefloat{figure}
\begin{document}

\begin{flushright}
kcl-mth-11-05\\
23 March 2011
\end{flushright}

\thispagestyle{empty}

\begin{center}
\vspace{14mm}
{\Large\bf 
Mean field theory for boundary Ising and\\[2mm] tricritical Ising models
}\\[2cm]
{\Large Philip Giokas}
\\[1cm]
{Department of Mathematics, King's College London,\\[2mm] Strand, London WC2R 2LS}
\\[2cm]

{\Large Abstract}
\\
\end{center}

\begin{quote}
Using the technique of mean field theory applied to the lattice boundary Ising and tricritical Ising models we provide a qualitative description of their boundary phase diagrams. We will show this is in agreement with the known picture from boundary conformal field theory and we shall compare our work with that of Cappelli et al and show how their analysis is not in accordance with the physical picture.  
\end{quote}
\newpage

\section{Introduction}
At criticality the  lattice Ising and tricritical Ising models become conformally invariant and one can represent their behaviour using the $M(3,4)$ and $M(4,5)$ conformal field theories as is the case for all higher critical Ising models. On introduction of a boundary there have been found different boundary states with specific conditions that maintain conformal invariance.  The simplest of these, and the ones we shall consider here, are the boundary states as found by John Cardy \cite{cardy1} known as the Cardy states. When considering the representative Ising and tricritical Ising lattice models, these Cardy states have physical interpretations in terms of the level occupancy for the boundary spins \cite{pearce}. These being $\{\pm1\}$ for the Ising model and $\{0,\pm1\}$ for the tricritical Ising model.  These boundary states have been analyzed in the literature and it is known that there are relevant boundary fields present at some of the boundaries and on perturbation by one or more of these relevant fields a renormalization group flow will take one to another of the conformally invariant boundary conditions.  For the Ising and tricritical Ising models under consideration here the details of the space of boundary RG flows has been found \cite{affleck} and the flows express irreversibility as one would expect from the g-theorem \cite{gtheorem}, which states there exists a non increasing g-function on the space of boundary RG flows analogous to the c-theorem \cite{ctheorem} in the bulk case.

The boundary phase diagram for the Ising models under consideration has already been examined by Cappelli et al \cite{cappelli} in which they used a Landau Ginzburg type polynomial boundary potential.  We however obtain a different picture based on a full mean field theory treatment and we will show when a Landau polynomial approximation is valid to the mean field equations and when it is not.  The main present motivation for this is to develop techniques that can then be applied to the more complex case of the space of RG flows in the case of conformal defects.  Furthermore as will be later elucidated not all of the flows in the boundary tricritical Ising model are integrable and our mean field description provides a qualitative picture of these flows.

The layout of the paper is as follows, in the second section we will first outline the Ising space of boundary conformal flows and then we will derive the mean field phase diagram for both the bulk and boundary.  In section three we will do exactly the same for the tricritical Ising model, in section four we will compare our picture with that of Cappelli et al and discuss when a Landau polynomial approximation is valid.  In section five we will draw our conclusions.

\section{The Ising model}
\subsection{Conformal boundary RG flows}
The Ising bulk conformal field theory comprises three symmetric holomorphic/anti\-holo\-morphic representations of the Virasoro algebra and Cardy has shown that there are thus three conformal boundary states. Two of these states are stable in that they have no relevant boundary fields and these correspond to the boundary spins being completely fixed in either of the $\{\pm1\}$ spin levels and these fixed boundary states are respectively denoted $(+)=(1,1)$ and $(-)=(2,1)$, where we have included the boundary states' Kac table index pairs.   Finally one has the degenerate boundary state $(d)=(1,2)$ where in the underlying lattice model fluctuations in the boundary spins take place equally across both $\{\pm1\}$ levels. This degenerate boundary state has one relevant field $\phi_{1,2}$ that the  boundary magnetic field $h_{b}$ couples to and switching on such a perturbation one induces an RG flow to either of the stable boundary states $(\pm)$ with the fixed point values being $h_{b}=\pm \infty$ respectively.  Once at these fixed points as there are no relevant fields there are no more RG flows and the flows thus respect irreversibility as proposed by g-theorem. We now illustrate the simple space of flows where the dark dots indicate stable boundary states and the white dots represent unstable boundary states.
\\
\begin{figure}[H]
	\centering
		\includegraphics{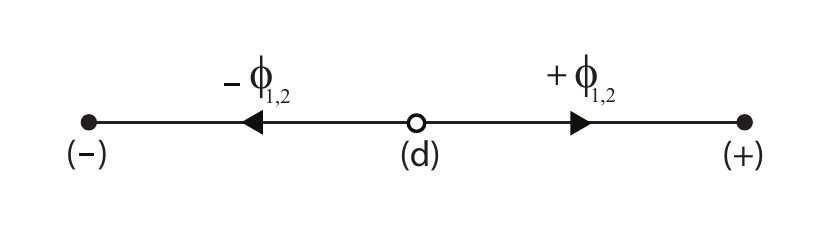}
		\caption{Conformal boundary RG flows for Ising model}
	\label{fig:bIsingRGflow}
\end{figure}
\subsection{The mean field bulk model}
The square lattice Ising model we take as representative of the M(3,4) minimal model is composed of spins $\sigma_{i,j}$ taking the values $\{\pm1\}$, $i$ represents the horizontal lattice direction and $j$ the vertical.  We define the lattice action to be
\begin{align}
H=\sum_{i,j}\left(-J\sigma_{i,j}(\sigma_{i+1,j}+\sigma_{i,j+1})-h\sigma_{i,j}\right)  \label{isaction}
\end{align}
Here $J$ is taken to be a positive (ferromagnetic) nearest-neighbour coupling and $h$ is an applied external magnetic field.  We utilise the mean field approximation as described in \cite{cardy2}, this consists of first re-expressing the spin degrees of freedom $\sigma$ as deviations $\delta\sigma=(\sigma-M)$ from their average value $M$ resulting in $\sigma=\delta\sigma+M$.  Then the nearest neighbour interaction terms become $\sigma\sigma'=\sigma M' + \sigma'M-MM'+\delta \sigma\delta \sigma'$ and it is the last term, the product of fluctuations that one drops in the mean field approximation resulting in
\begin{align}
\sigma\sigma'=\sigma M' + \sigma'M-MM'  \label{mfapprox}
\end{align}
On entering this approximation into eq (\ref{isaction}) one obtains the mean field action
\begin{align}
H_{mf}=\sum_{i,j}\left(-J\sigma_{i,j}\widetilde{M}_{i,j}+\frac{J}{2}M_{i,j}\widetilde{M}_{i,j}-h\sigma_{i,j}\right) 
\end{align}
Here $M_{i,j}$ denotes the average magnetization at the lattice site $(i,j)$ and $\widetilde{M}_{i,j}$ denotes the sum of nearest neighbour magnetizations of the lattice site $(i,j)$ ie.
\begin{align} 
\widetilde{M}_{i,j}=M_{i,j+1}+M_{i,j-1}+M_{i+1,j}+M_{i-1,j} 
\end{align}
From this we can deduce the mean field partition function 
\begin{align}
Z_{mf}=\sum_{\left\{\sigma\right\}}\exp(-\beta H_{mf})=\prod_{i,j}\exp(-\frac{\beta J}{2}M_{i,j}\widetilde{M}_{i,j})\; 2\cosh(\beta(J\widetilde{M}_{i,j}+h))\label{ismfpart}
\end{align}
where the sum over $\{\sigma\}$ is taken over all lattice spin configurations and $\beta=1/T$ where $T$ is temperature. The reason that the mean field partition function can be evaluated is that the sum over the spins decouple meaning that
\begin{align}
\sum_{\{\sigma_{n}\}}\exp(-K_{n}\sigma_{n})=\prod_{n}\sum_{\sigma_{n}=\pm 1}\exp(-K_{n}\sigma_{n})=\prod_{n}2\cosh(K_{n}).
\end{align}
From equation (\ref{ismfpart}) we can obtain the mean field free energy 
\begin{align}
f_{mf}=-\frac{1}{\beta}\log(Z_{mf})=\frac{J}{2}\sum_{i,j}{M_{i,j}\widetilde{M}_{i,j}}-\frac{1}{\beta}\log( 2\cosh(\beta(J\widetilde{M}_{i,j}+h))).
\end{align}
Minimizing the free energy with respect to the magnetizations $M_{i,j}$ results in the mean field consistency equations for the site magnetizations
\begin{align}
M_{i,j}=\tanh(\beta(J\widetilde{M}_{i,j}+h))\label{ismag}  
\end{align}
\subsubsection{Zero bulk fluctuation limits}
Before going on to study the nature of any phase transitions in the bulk we can use the previous magnetization equation (\ref{ismag}) to deduce the regions of  the phase diagram that correspond to all the bulk spins being completely aligned in either of the $\{\pm1\}$ levels.  Our first step is to take the previous bulk magnetization equation (\ref{ismag}) and impose a flat magnetization profile as is expected in the bulk $M(x,y)=\bar{m}$. From this we can deduce the level occupancies $p_{\pm}$ that tell us the probability of finding a given spin in either of the $\{\pm\}$ states from the relation $p_{\pm}=(1\pm\bar{m})/2$ and  this results in 
\begin{align}
p_{\pm}=\frac{1}{2}(1\pm\mathrm{arctanh}(\beta(4J\bar{m}+h_{b})))
\end{align}
On letting $\beta h$ tend to infinite limits $(\beta h\rightarrow \pm \infty)$ the $\bar{m}$ term becomes insignificant and we can deduce the regions
\begin{align}
(\beta h \rightarrow \pm  \infty) \Rightarrow p_{\pm}\rightarrow 1.
\end{align}
These results are exact because far from any phase transitions mean field theory becomes exact.  The reason for this is that going back to our original unapproximated Hamiltonian (\ref{isaction}) and on taking the limits of $h\rightarrow\pm\infty$ the unapproximated $h\sigma$ terms will completely dominate the nearest neighbour terms $-J\sigma\sigma'$ that are finite on account of $J$ being finite.  One can deduce that if $p_{\pm}\rightarrow1$ exactly in the limits $\beta h\rightarrow\pm\infty$ then any phase transition from the $+1$ level to the $-1$ level must take place for finite $\beta h$.
\subsubsection{Bulk phase transition}
Having established the stable regions where all the spins are aligned completely in either orientation we can now go on to study the nature of the phase transition between the levels.  Going back to our original mean field equation (\ref{ismag}) we want to consider the magnetization to be a smoothly varying field living on the plane and thus we now re express the site magnetizations in terms of a continuous function $M_{i,j}=M(ia,ja)$ where $a$ is the lattice spacing.  Next we Taylor expand the $\widetilde{M}_{i,j}$ term to second order about the central coordinate $(ia,ja)$.  This then results in a differential continuum differential equation that approximates the bulk mean field equations (\ref{ismag})
\begin{align}
M=\tanh(\beta(4JM+a^2J\nabla^{2}M+h)).
\end{align}
This can be  inverted to  give the second order differential equation
\begin{align}
a^2\beta J\nabla^{2}M=\mathrm{arctanh}(M)-4\beta JM-\beta h.  \label{isbulkde}
\end{align}
The rhs of the previous equation can be Taylor expanded for small $M$ to give the second order equation
as obtained in \cite{cardy1}
\begin{align}
\beta a^{2}J\nabla^{2}M=(1-4\beta J)M+\frac{1}{3}M^{3}-\beta h. \label{cardyeq}
\end{align}
In the bulk we impose a flat field profile with constant value denoted $M(x,y)=\bar{m}\;\forall x,y$ this in turn implies the second derivative term goes to zero and we can rewrite the previous equation as
\begin{align}
0=-\beta h+\left(1-4\beta J\right)\bar{m}-\frac{1}{3}\bar{m}^{3} . \label{isbulkcriteq}
\end{align}
The condition for a critical point is a  point of inflection on the curve of $h$ against $\bar{m}$ or $\partial_{\bar{m}}h=\partial_{\bar{m}}^{2}h=0$; this condition is achieved at $\bar{m}=0$, with $h=0$ and $\beta J=1/4$.  The expansion of the rhs of equation (\ref{isbulkde}) that resulted in equation (\ref{cardyeq}) we will refer to in the rest of the paper as a Landau approximation.  The reason being that these equations are the same as the ones which would have been obtained from the Landau-Ginzburg effective action
\begin{align}
S=\int \mathrm{d}x\mathrm{d}y\left(\frac{1}{2}(\nabla M(x,y))^{2}-\beta h M(x,y)+\frac{1}{2}(1-4\beta J)M(x,y)^{2}+\frac{1}{12}M(x,y)^{4}\right). \nonumber \\ \label{island}
\end{align}
The integrand in equation (\ref{island}) would be obtained by Taylor expanding the free energy for small and slowly varying $M(x,y)$, in such an expansion the coefficients of the different powers of $M(x,y)$ have a functional dependence on the theory's parameters and it is the behaviour of these coefficients that dictates the nature of any critical points and phase structure around them.  In the Landau theory, after making such an expansion one truncates the expansion at the smallest power of $M$ whose coefficient stays positive under changes of the theory's parameters.   The key point is that the order parameter $M$ should be small for such an expansion to be valid.
\subsubsection{Mean field boundary model}
Now we consider the lattice Ising model on the half plane and with the boundary running parallel to the vertical $j$ direction and with the boundary's horizontal coordinate set at $i=0$. The boundary conditions are set by a boundary magnetic field $h_{b}$ acting on the boundary spins $\sigma_{0,j}$ and a ferromagnetic nearest neighbour boundary coupling $J_{b}$ that acts between the boundary spins, finally the boundary spins interact with the  bulk spins via the bulk interaction $J$.  These conditions are reflected in the following half plane Ising lattice action
\begin{align}
H=\sum_{i>0,j}\left(-J \sigma_{i,j}(\sigma_{i+1,j}+\sigma_{i,j+1})-h\sigma_{i,j}\right)+\sum_{j}\left(-J \sigma_{0,j}\sigma_{1,j}-J_{b}\sigma_{0,j}\sigma_{0,j+1})-h_{b}\sigma_{0,j}\right).
\end{align}
We use the same mean field spin interaction approximation equation (\ref{mfapprox}) as before for deriving the mean field partition function and further we assume that the only variation in magnetization will occur in the horizontal direction perpendicular to the boundary, meaning that  $M_{i,j}=M_{i,j-1}=M_{i,j+1}\; \forall\; i,j$ and this results in a free energy per lattice row 
\begin{align}
f_{mf}=&J_{b}M^{2}_{0}+JM_{0}M_{1}-\frac{1}{\beta}\log(2\cosh(\beta(2J_{b}M_{0}+JM_{1})+h_{b}))  \\
&+\sum_{i>0}\left(JM_{i}M_{i+1}-\frac{1}{\beta}\log(2\cosh(\beta(J(M_{i-1}+2M_{i}+M_{i+1})+h)))\right). \nonumber
\end{align}
Minimizing this with respect to the magnetization, $\partial f_{mf}/\partial M_{i}=0\; \forall i$, we obtain the mean field magnetization equations.
\begin{align}
&M_{0}=\tanh(\beta(2J_{b}M_{0}+JM_{1})+h_{b}) \label{isbdymag}  \\
&M_{i}=\tanh(\beta(J(M_{i-1}+2M_{i}+M_{i+1})+h))\;\;\;(i>0) \label{isbulkmag} 
\end{align}
The following steps for obtaining a boundary equation for studying the boundary phase diagram are our own extension to the standard mean field techniques used in \cite{cardy2}.  Firstly, as in the bulk case we again re-express the discrete magnetizations $M_{i}$ in terms of a continuous function $M_{i}=M(ia)$, where $a$ is the lattice spacing and it is understood the magnetization function $M(x)$ only depends on the horizontal plane coordinate $x$ and thus the boundary magnetization value is simply $M(0)$.  The boundary equation (\ref{isbdymag}) then becomes, on inversion and rearrangement,
\begin{align}
0=\mathrm{arctanh}(M(0))-2\beta J_{b}M(0)-\beta  JM(a)-\beta h_{b}.\label{isbdyinit}
\end{align}
It is this equation that we will use to obtain the boundary phase diagram, as it relates the boundary field $h_{b}$ with magnetization $M(0)$ and any critical points are defined in terms of derivatives of $h_{b}$ with respect to $M(0)$.  The magnetization one lattice spacing from the boundary, $M(a)$, in equation (\ref{isbdyinit}) will depend on the boundary magnetization $M(0)$ via the bulk equations (\ref{isbulkmag}) and thus we need to obtain this dependence in order to end up with an expression purely in terms of $h_{b}$ and $M(0)$.  The strategy we shall employ will be the following: we first take the bulk equations (\ref{isbulkmag}) in their continuum form, $M_{i}=M(ia)$ and we Taylor expand the magnetization $M_{i\pm1}=M(ia\pm a)$ term about $(ia)$ to second order. On inversion and rearrangement we obtain the same differential equation as we did in the bulk equation (\ref{isbulkde}) 
\begin{align}
a^2\beta J\partial^{2}M=\mathrm{arctanh}(M)-4\beta JM-\beta h.\label{isprofile}
\end{align}
However, as there is by construction no variation in the vertical direction, we have in this case $\nabla^{2}=\partial^{2}$ where $\partial$ will be used from now on to denote the  derivative with respect to the horizontal direction.  We take it that this differential equation exactly defines the form that $M(x)$ will take and furthermore that it not only holds in the bulk $(x>0)$ as it came from the bulk magnetization equations but also at the boundary $(x=0)$ in order that we obtain a smoothly varying profile. As it is a second order equation we need to impose two boundary conditions in order to obtain the profile $M(x)$. The first  boundary condition will come from the fact that we impose a flat profile in the bulk limit $x\rightarrow \infty$, the value of the magnetization in this bulk limit we denote $\lim_{x\rightarrow \infty}M(x)=M_{\infty}$.  This will entail that the second derivative term in our bulk profile differential equation will tend to zero in the bulk limit and we will be left with the bulk limit equation
\begin{align}
0=\mathrm{arctanh}(M_{\infty})-4\beta JM_{\infty}-\beta h 
\end{align}
This is the same as the bulk theory equation (\ref{isbulkcriteq}) obtained in the previous section  and thus we find the same critical parameters of $h=0$ and $\beta J=1/4$ with the result $M_{\infty}=0$.  As we are looking to hold the bulk critical while varying the boundary parameters, we shall insert these critical bulk values into our profile differential equation (\ref{isprofile}) in order that we obtain a critical bulk profile equation
\begin{align}
a^2\partial^{2}M=4(\mathrm{arctanh}(M)-M).
\end{align}
As previously mentioned, one of the boundary conditions will be the critical bulk limit $M_{\infty}=0$ and the second will be the value of the boundary magnetization $M(0)$. We can integrate this second order equation once including the bulk limit boundary condition $M_{\infty}=0$ to obtain a first order critical profile equation with only one boundary condition the boundary magnetization $M(0)$ to be set. We use the result
\begin{align}
\frac{1}{2}(\partial M)^2=\int_{0}^{M}\partial^{2}M\;\; \mathrm{d}M
\end{align}
to find
\begin{align}
a\partial M(x)=\pm2\sqrt{2M\mathrm{arctanh}(M)+\log(1-M^{2})-M^{2}}. \label{isfirstode}
\end{align} 
As shown in Cappelli et al \cite{cappelli} the correct choice of sign in equation (\ref{isfirstode}) is positive if the value of the magnetization is negative as the magnetization should monotonically increase to the bulk limit value of zero and vice versa.  So this first order equation defines the form of $M(x)$ given a boundary value of $M(0)$ and this means that the $M(a)$ term in the boundary equation can be expressed in terms of $M(0)$ in the following way.  Given that we have the values of the first and second derivatives at a point we can use them together to obtain all higher order derivatives using the chain rule 
\begin{align}
\partial^{n}M=\partial_{M}(\partial^{n-1}M) \partial M .
\end{align}
And with knowledge of all the derivatives we can write down a series expansion for $M(a)$ about the boundary resulting in an expression purely in terms of the boundary magnetization $M(0)$ in the form of an expansion $B_{e}(M(0))$
\begin{align}
M(a)=B_{e}(M(0))=\sum_{n=0}^{\infty}{\frac{a^{n}\partial^{n}M(0)}{n!}}. \label{Bexp} 
\end{align}
Even though we can never consider the whole expansion we will  find that at critical points where $M(0)=0$ we will only need to consider the behaviour of the first two terms (the behaviour of the rest follows by induction) and when $M(0)\neq  0$ as occurs when considering the tricritical model we can suitably truncate the expansion to a degree that does not appreciably affect any numerical answers.  With this expansion and on letting $M(0)=m$ our original boundary equation becomes 
\begin{align}
0=\mathrm{arctanh}(m)-2\beta J_{b}m-\frac{1}{4}B_{e}(m)-\beta h_{b}. \label{isbdymaster} 
\end{align}
This is an expression purely between $h_{b}$ and $m$ that we can use to study the boundary phase diagram.  Before we go on to obtain the boundary phase diagram we will point out that we cannot use the above equation to plot $m$ vs $h_{b}$ across the whole range $-1\leq m \leq 1$ as the expansion $B_{e}(m)$ only converges for the range $|m| \leq 0.6$.  To overcome this we will obtain a numerical form of the function $M(a)=B_{e}(m)$ denoted $M(a)=B_{n}(m)$ and will use it when making plots of the full mean field equations across the whole range of boundary magnetization.  The way we obtain this function is to numerically solve our critical first order profile equation (\ref{isfirstode}) a large number of times each time with a different value of boundary condition $m$ taken evenly from the full range $-1\leq m \leq 1$.  Then for each of these different $m$ dependent numerical solutions we read off the corresponding value of $M(a)$ and tabulate the results in a list of $M(a)$ vs $m$ and then we use Mathematica to define a interpolative numerical function $M(a)=B_{n}(m)$ from said table.  
\\
\begin{figure}[H]
	\centering
		\includegraphics{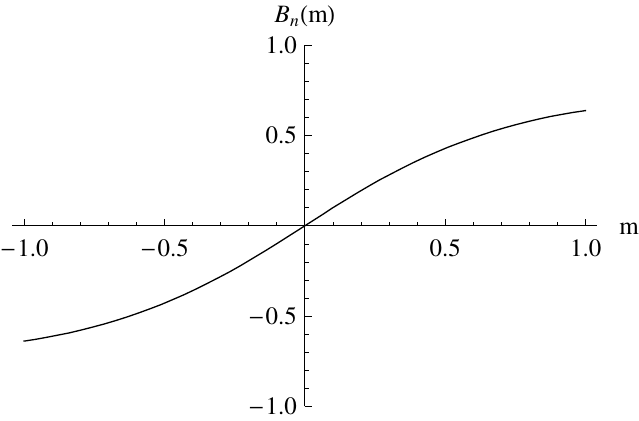}
		\caption{Plot of numerical function $M(a)=B_{n}(m)$ that gives the magnetization a lattice spacing from the boundary in terms of the magnetization at the boundary}
	\label{fig:isingbn}
\end{figure}
With this we obtain a numerical boundary equation that can be used to plot the full mean field behaviour of $m(h_{b})$
\begin{align}
0=\mathrm{arctanh}(m)-2\beta J_{b}m-\frac{1}{4}B_{n}(m)-\beta h_{b}.  \label{isbdyeqnum} 
\end{align}
However it  is the expansion form of the equation (\ref{isbdymaster}) that we will use to deduce the existence and nature of any  critical points as will follow.  First though we examine the regions of  the  phase diagram where there are no boundary fluctuations.
\subsubsection{Zero boundary fluctuation limits}
In analogy with the bulk there are limits of the boundary phase diagrams that  correspond to fixing all the boundary spins either of the $\{\pm1\}$ levels.  In this case we need the boundary occupancy probabilities $p_{\pm}$ which we obtain from the same procedure as in the bulk to be
\begin{align}
p_{\pm}=\frac{1}{2}(1\pm\mathrm{arctanh}(\beta(K_{0}+h_{b}))) \label{isbdyprob}
\end{align}
where $K_{0}=2J_{b}M(0)+JM(a)$ and this term is always small and finite and thus as in the bulk it is completely dominated by any large value of $h_{b}$ and we find the regions
\begin{align}
(\beta h_{b} \rightarrow \pm  \infty) \Rightarrow p_{\pm}\rightarrow 1 
\end{align}
and we can identify these respective limits $\beta h_{b}\rightarrow \pm\infty$ with the stable Cardy states $(\pm)$ that correspond to aligning all the boundary spins in one orientation.
\subsubsection{Boundary phase transition}
\begin{figure}[H]
    \centering
   (a)
    {
        \includegraphics{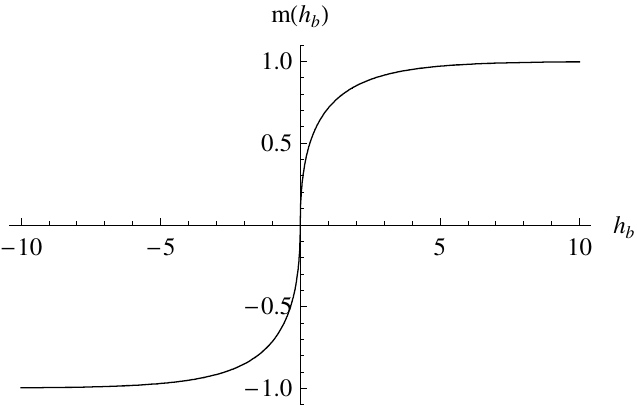}
        \label{fig:second_sub}
     }
  (b)
    {
        \includegraphics{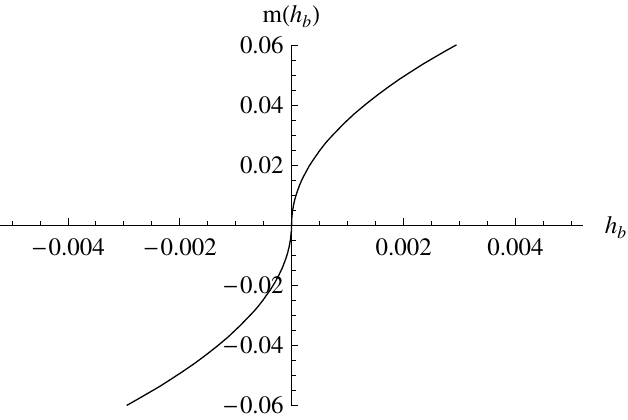}
        \label{fig:third_sub}
      }
    \caption{Plots of $m(h_{b})$ using (a) the full mean field equations and (b) the Landau approximated version of the mean field equations for a smaller range of $m$ as the Landau approximation only holds for small values of m}
    \label{fig:iscritmofh}
\end{figure}

As in the bulk, we can expect a phase transition at $h_{b}=m=0$ that separates the stable limits considered before.  As a starting point we take the boundary equation (\ref{isbdymaster}) obtained before
\begin{align}
0=\mathrm{arctanh}(m)-2\beta J_{b}m-\frac{1}{4}B_{e}(m)-\beta h_{b}.\nonumber
\end{align}
We cannot expand the rhs of the equation (\ref{isbdymaster}) in powers of $m$ as is normally the case due to the behaviour of the odd terms in $B_{e}(m)$ coming from taking different signs of the square root in our first derivative term.  We can, though expand, in powers involving $\left|m\right|$ resulting in
\begin{align}
0=-\beta h_{b}+(\frac{3}{4}-2\beta J_{b})m+\frac{1}{2\sqrt{6}}m\left|m\right|+O(m^{3}\left|m\right|). \label{isbdylan}
\end{align}
We obtain a phase transition for $h_{b}=m=0$ when the linear term in equation (\ref{isbdylan}) vanishes at $\beta J_{b}=3/8$ and this entails $\partial_{m}h_{b}=0$ as is required for a critical point. However, normally, one requires $\partial^{2}_{m}h_{b}=0$ as well, but the second derivative is not defined at $m=0$  in the same way that the first derivative of $\left|m\right|$ is not defined at $m=0$, though graphically it at least looks like we have a point of inflection.  This critical point we identify with the $(d)$ boundary state from the known conformal picture.  In figure (\ref{fig:iscritmofh}) we plot the phase transition using both the numerical form of the full mean field equation and also the polynomial Landau approximation where it should be noted that we have to set the bulk interaction $J$ and we take it to be $J=1$ in the all the plots in this paper.
The full mean field plot qualitatively illustrates the picture from the conformal field theory with RG flows taking one from the $(d)$ state at $h_{b}=0$ to either of the stable states $(\pm)$ with $m=\pm1$.  From now on in we shall only plot phase transitions using the full mean field equations as the Landau approximation plots just elucidate the behaviour of the full mean field plots for small values of magnetization ie they don't add any information.
\subsubsection{Superpotential}
\begin{figure}[H]
    \centering
   
    (a)
    {
        \includegraphics{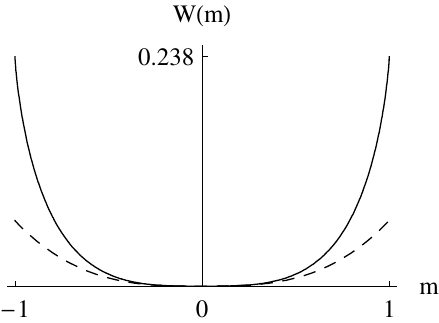}
        \label{fig:second_sub}
     } \;\;\;\;\;\;\;\;\;\;\;\;\;\;\;\;
    (b)
    {
        \includegraphics{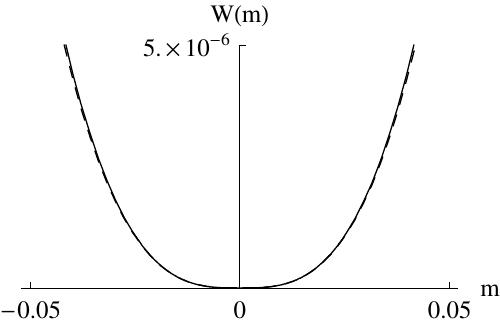}
        \label{fig:third_sub}
      }
    \caption{Plots of superpotential $W(m)$ for large (a) and small (b) $m$, the dashed plot corresponds to the Landau polynomial approximation and the full plot corresponds to the full mean field superpotential, note that in figure (b) plots are coincident near origin.}
    \label{fig:iscritsup}
\end{figure}

\begin{figure}[H]
	\centering
		\includegraphics{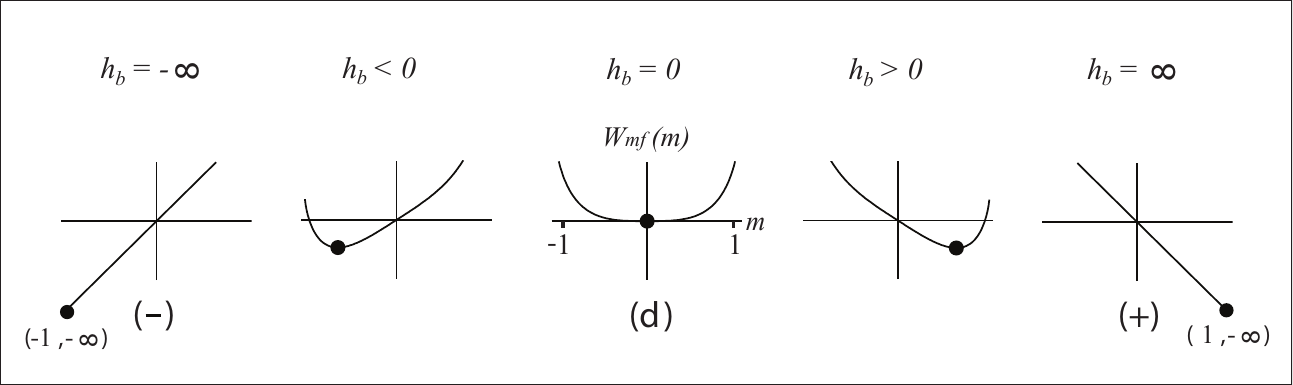}
		\caption{Mean field Ising boundary phase diagram}
	\label{fig:isingbph}
\end{figure}

It is sometimes useful to consider the boundary equation the master boundary equation (\ref{isbdymaster}) as coming from the minimization of a function of the boundary magnetization $m$.  This was first considered by Cappelli et al in \cite{cappelli} where they called this function the "`superpotential"'.  Accordingly we define our superpotential $W(m)$ such that the condition $\partial_{m}W=0$ gives rise to our master boundary equation (\ref{isbdymaster}) .  This is only possible exactly when taking the Landau polynomial approximation of the equations as the terms in $B_{e}(m)$ are not amenable to exact integration.  However we can numerically obtain a superpotential $W(m)$ of the full mean field equation as $B_{n}(m)$ can be integrated numerically, this allows us to make plots of the superpotential across the whole range $-1\leq m \leq 1$ and also when a Landau type expansion is not possible as occurs at some critical points of the tricritical model.  Taking the Landau form of our boundary equations and integrating we obtain
\begin{align}
W_{Lan}(m)=-\beta h_{b}m+(\frac{3}{8}-\beta J_{b})m^{2}+\frac{1}{6\sqrt{6}}m^{2}\left|m\right| \label{issuplan}
\end{align}
if we tune $\beta J_{b}$ to its critical value of $3/8$ then this expression is equivalent to the superpotential expression obtained by Cappelli
\begin{align}
W_{cap}(m)=a_{0}m+\frac{1}{3}m^{2}\left|m\right|
\end{align}
with $-\beta h_{b}=a_{0}$.  As stated before we can also obtain a mean field superpotential $W_{mf}(m)$ taken by integrating the full mean field boundary equation with the numerical $B_{n}(m)$ integrated numerically, this results in
\begin{align}
W_{mf}(m)=m\mathrm{arctanh}(m)+\frac{1}{2}\log(1-m^{2})-\beta J_{b}m^{2}-\frac{1}{4}\int B_{n}(m)\mathrm{d}m-\beta h_{b}m
\end{align}
In figure (\ref{fig:iscritsup}) we plot the superpotential in both full mean field form $W_{mf}(m)$ and in the polynomial approximation.
as was the case for the phase transition plots we shall only plot the full mean field superpotential from now on. It should be noted that in subfigure (a) figure (\ref{fig:iscritsup}) the vertical axis increment of 0.238 is chosen as $W_{mf}(\pm1)=0.238\;\mathrm{3sf}$. Outside of the range \newline $-1\leq m\leq1$ one finds that $W_{mf}(m)$ becomes complex and unphysical and therefore the mean field picture naturally retains the magnetization in this range, however the Landau approximated superpotential (\ref{issuplan}) can be plotted across all values of $m$ and does not have this retantion property. In fact all the mean field superpotentials considered in this paper have this retention property. An advantage of this is that it allows one to deduce the level probabilities (\ref{isbdyprob}) for large $m$ whereas in the Landau picture where there is no such retention one loses this information.  In figure (\ref{fig:isingbph}) we illustrate the boundary Ising phase diagram using our full mean field superpotential.  

\section{The tricritical Ising model}
\subsection{Conformal RG boundary flows}
\begin{figure}[H]
	\centering
		\includegraphics{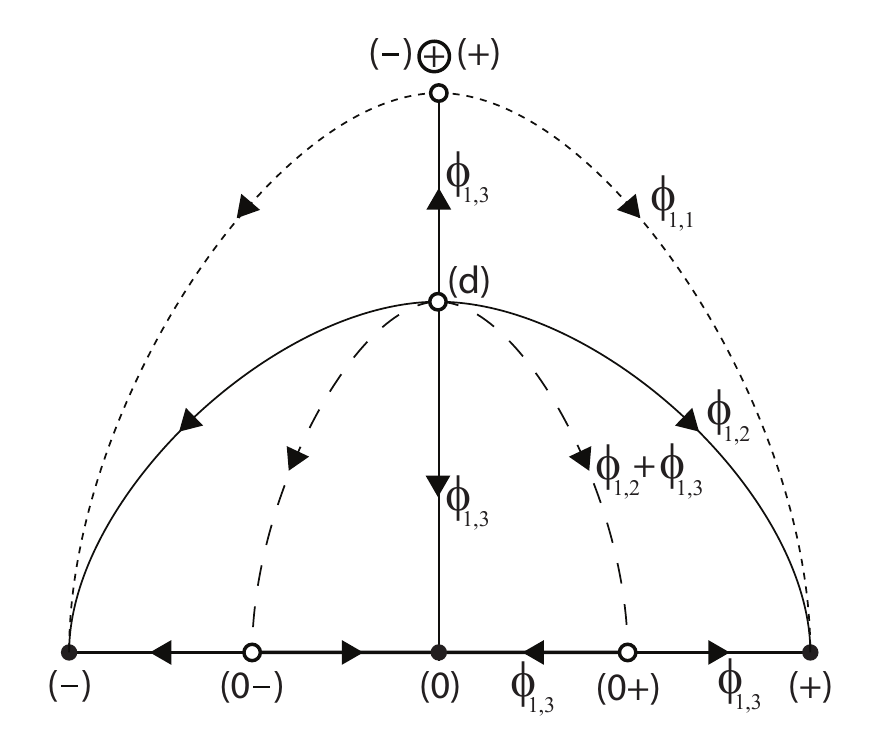}
		\caption{Space of boundary RG flows from Ref. \cite{affleck}, finely dashed lines represent 1st order transitions, coarsely dashed represent non integrable flows}
	\label{fig:btriRGflow}
\end{figure}

The tricritical bulk theory is described by the $M(4,5)$ minimal model and has six holomorphic/antiholomorphic representations of the Virasoro algebra and thus there are six corresponding Cardy states when one introduces a boundary.  As in the case of the Ising model these can be related to the conditions on the boundary spins of the representative tricritical ising lattice model where the spins can take the values $\{\pm1,0\}$ which correspond to spin up/down and spin zero or no spin respectively.  There are  stable boundary states that have no relevant boundary fields, these correspond to fixing all the boundary spins in a given orientation.  They are denoted as follows where we also include their corresponding Kac table indexes: $(+)=(1,1),\;(-)=(3,1),\;(0)=(2,1)$.  We also have two mixed boundary states $(0+)=(1,2)$ and $(0-)=(1,3)$ that correspond to allowing the boundary spins to only fluctuate between the levels $\{0,+1\}$ and $\{0,-1\}$ respectively.  These states have one relevant boundary field $\phi_{1,3}$ and perturbing with this field will induce an RG flow to one of the given stable states as will be shown in the next figure. The last of the Cardy boundary states is the degenerate boundary state $(d)$ that corresponds in the lattice model to letting the boundary spins fluctuate across all levels.  It has two relevant boundary fields $\phi_{1,2}$ and $\phi_{1,3}$, on perturbations by these fields one obtains an RG flow to the other Cardy states as is elucidated in figure (\ref{fig:btriRGflow})
\\

As one can see from figure (\ref{fig:btriRGflow}) the mixed flows taking one to either of the $(\pm0)$ states are not integrable and hence our following mean field boundary phase diagram will give a qualititive picture of them.  One can also consider a superposition of the stable boundary states $(+)\oplus(-)$ as first show in \cite{reck}.  This superposition has one relevant boundary field given by the second identity field, on perturbation by this field one obtains a first order transition (a discontinuous jump in magnetization) to either of the stable states $(+)$ or $(-)$.  In the next subsection we shall apply mean field theory to the tricritical Ising lattice model in order to obtain a qualitative picture of the boundary phase diagram and then we shall see how this compares with the conformal picture just described.
\subsection{The mean field bulk model}
Now we consider the tricritical Ising model where we allow the spins $\sigma$ to take the value $\{0\}$ as well as $\{\pm 1\}$, alternatively the $\{0\}$ state can be viewed as a spin vacancy.  The square lattice action we shall take is 
\begin{align}
H=\sum_{i,j}\left(-J\sigma_{i,j}(\sigma_{i+1,j}+\sigma_{i,j+1})-h \sigma_{i,j}+\Delta\sigma_{i,j}^{2}\right).
\end{align}
The parameter $\Delta$ is known as the vacancy density and increasing it in the positive  direction increases the number of vacancies and increasing it in the negative direction decreases the number of vacancies.  The parameters $J$ and $h$ are as before a ferromagnetic nearest neighbour coupling and an applied magnetic field. On applying the same mean field substitution (\ref{mfapprox}) of the nearest neighbour terms in the action we obtain the mean field action
\begin{align}
H_{mf}=\sum_{i,j}\left(-J\sigma_{i,j}\widetilde{M}_{i,j}+\frac{J}{2}M_{i,j}\widetilde{M}_{i,j}-h\sigma_{i,j}+\Delta\sigma_{i,j}^{2}\right).
\end{align}
From this we obtain the mean field partition function 
\begin{align}
Z_{mf}=\prod_{i,j}\exp(-\frac{\beta J}{2}M_{i,j}\widetilde{M}_{i,j})\; (1+2y\cosh(\beta(J\widetilde{M}_{i,j}+h)))
\end{align}
where $y=e^{-\beta \Delta}$ and from this follows the free energy
\begin{align}
f_{mf}=\frac{J}{2}\sum_{i,j}{M_{i,j}\widetilde{M}_{i,j}}-\frac{1}{\beta}\log(1+2y\cosh(\beta(J\widetilde{M}_{i,j}+h))).
\end{align}
Minimizing the free energy with respect to the magnetizations $M_{i,j}$ results in the mean field consistency equations for the magnetization
\begin{align}
M_{i,j}=\frac{2y\sinh(\beta(J\widetilde{M}_{i,j}+h))}{1+2y\cosh(\beta(J\widetilde{M}_{i,j}+h))}\label{tribulkmag}
\end{align}
before using this equation to study the bulk phase transitions we will use it to study regions of the  phase diagram where all the spins completely occupy one of the $\{0,\pm1\}$ levels ie where there are zero fluctuations.
\subsubsection{Zero bulk fluctuation regions}
In order to deduce the regions we need the level occupancies $p_{\pm}$ and $p_{0}$ obtainable from the relations $p_{0}=1-\left\langle\sigma^{2}\right\rangle$ and $p_{\pm}=(\left\langle\sigma^{2}\right\rangle\pm\left\langle\sigma\right\rangle)/2$. We can obtain the expectation value of $\left\langle \sigma^{2}\right\rangle$ by differentiating $f_{mf}$ wrt to $\Delta$ and we obtain it as the same expression for the magnetization expect with the sinh turning to a cosh.  In the bulk theory we impose a flat magnetization profile and this means that $M_{i,j}=\bar{m}\; \forall i,j$ where $\bar{m}$ is the average bulk magnetization.  On inserting this into our magnetization and $\left\langle \sigma^{2}\right\rangle$ equations and following the previously mentioned relations we obtain the level occupancies
\begin{align}
p_{+}=\frac{e^{\beta(4J\bar{m}-(\Delta-h))}}{1+e^{\beta(4J\bar{m}-(\Delta-h))}+e^{\beta(-4J\bar{m}-(\Delta+h))}} \\ \nonumber \\  
p_{-}=\frac{e^{\beta(-4J\bar{m}-(\Delta+h))}}{1+e^{\beta(4J\bar{m}-(\Delta-h))}+e^{\beta(-4J\bar{m}-(\Delta+h))}}\\ \nonumber \\ 
p_{0}=\frac{1}{1+e^{\beta(4J\bar{m}-(\Delta-h))}+e^{\beta(-4J\bar{m}-(\Delta+h))}}.
\end{align}  
Using these expressions we deduce the zero fluctuation regions as
\begin{align}
&\beta(\Delta-h)\rightarrow\infty\;\mathrm{and}\;\beta(\Delta+h)\rightarrow\infty\;         \Rightarrow p_{0}\rightarrow 1  \\
&\beta(\Delta-h)\rightarrow -\infty\;\mathrm{and}\;\beta h\rightarrow \infty \;\;\;\;\;\;\;\;\Rightarrow p_{+}\rightarrow 1  \\
&\beta(\Delta+h)\rightarrow -\infty\;\mathrm{and}\;\beta h\rightarrow -\infty\;\;\;\;\;\Rightarrow p_{-}\rightarrow 1 
\end{align}
We will analyze these regions in more detail in the boundary case where the analysis is essentially the same as in the bulk case.
\subsubsection{Bulk phase transitions}
Now we will examine the phase transitions present in the bulk model and for this we need to consider our previous magnetization equation (\ref{tribulkmag}). This can be inverted, the $\widetilde{M}_{i,j}$ Taylor expanded and the $M_{i,j}$ can be recast as $M(ia,ja)$ as before resulting in 
\begin{align}
&a^{2}\beta J\nabla^{2}M=\mathrm{F}(M,y)-4\beta JM-\beta h \label{nctribulkde}
\end{align}
where the function $F(M,y)$ is defined as follows 
\begin{align}
\mathrm{F}(M,y)=\log\left(\frac{1}{2y}\left(\frac{M}{1+M}+\sqrt{\left(\frac{M}{M+1}\right)^{2}+4y^{2}\left(\frac{M+1}{M-1}\right)}\right)\right).
\end{align}
The magnetization profile $M=M(x,y)$ must obey the differential equation (\ref{nctribulkde}) in the bulk, however as there is no variation in the bulk we take the profile to be again constant $M(x,y)=\bar{m}\;\forall x,y$ this then gives
\begin{align}
0=&\mathrm{F}(\bar{m},y)-4\beta J\bar{m}-\beta h.
\end{align}
This can be expanded in a Landau approximation for small $\bar{m}$ giving
\begin{align}
0=&-\beta h+ (\frac{1+2y}{2y}-4\beta J)\bar{m}+\frac{(4y-1)(1+2y)^{2}}{48y^{3}}\bar{m}^{3}  \\ &+\frac{(3-18y+32y^{2})(1+2y)^3}{1280y^{5}}\bar{m}^{5}+\mathrm{O}(\bar{m}^{6}) \nonumber
\end{align}
This is  critical if $\partial_{\bar{m}}h=\partial_{\bar{m}}^{2}h=0$, which is equivalent to the linear and constant terms going to zero in the above ie $h=0$ and $\beta J=(1+2y)/8y$.  Furthermore it is tricritical if $\partial_{\bar{m}}^{3}h=0$ which corresponds to the cubic term vanishing as well which happens for 
\begin{align}
h=0,\;y=\frac{1}{4},\; \beta J=\frac{3}{4}
\end{align}
Thus we have a tricritical point at these coordinates and from here increasing $y$, which corresponds to decreasing $\Delta$ takes one along a line of critical points and in the limit $y\rightarrow \infty \Rightarrow \Delta \rightarrow -\infty$ we recover the Ising bulk mean field equation and in turn the critical Ising phase transition.  Furthermore we can find two more lines of critical points adjoining the tricritical point, in fact it is the joining of the three lines of critical points that gives the tricritical point its name.  These two lines of critical points occur roughly along the lines $(h=\pm \Delta)$ and we shall only consider the limits $(h\rightarrow \pm \infty, \Delta \rightarrow \infty \Rightarrow y\rightarrow 0)$ as it will be these limits that are of interest when we will consider the boundary theory.  On taking these limits our bulk equation takes the form
\begin{align}
0=2\mathrm{arctanh}(2\bar{m}\mp1)-4\beta J\bar{m} -\beta(h\mp \Delta).
\end{align}
These equations cannot be expanded for small $\bar{m}$ as the arctanh term diverges at $m=0$.  However we can still solve it exactly and one finds critical points when $\bar{m}=\pm 1/2$, $\beta J=1$ and $\beta(h\mp \Delta)=\mp2$.  Having found the critical values of $\bar{m}=\pm 1/2$ we can Landau expand around these critical points by defining $m_{\pm}=\bar{m}\mp 1/2$
\begin{align}
0=-\beta(h\mp \Delta\mp 2 J)+(4-4\beta J)m_{\pm}+\frac{16}{3}m_{\pm}^{3}
\end{align}
\subsection{The mean field boundary model}
As in the case of the Ising model we consider the tricritical model defined on the infinite half plane with lattice action
\begin{align}
H=&\sum_{i>0,j}\left(-J \sigma_{i,j}(\sigma_{i+1,j}+\sigma_{i,j+1})-h\sigma_{i,j}+\Delta\sigma_{i,j}^{2}\right)\label{tribdyaction} \\
&+\sum_{j}\left(-J \sigma_{0,j}\sigma_{1,j}-J_{b}\sigma_{0,j}\sigma_{0,j+1}-h_{b}\sigma_{0,j}+\Delta_{b}\sigma_{0,j}^{2}\right)  \nonumber
\end{align}
here the boundary parameters are denoted with subscript $b$. We assume the magnetization only varies perpendicular to the boundary and on applying the mean field approximation we obtain the free energy 
\begin{align}
f_{mf}=&J_{b}M^{2}_{0}+JM_{0}M_{1}-\frac{1}{\beta}\log(1+2y_{b}\cosh(\beta(2J_{b}M_{0}+JM_{1})+h_{b}))  \\ 
&+ \sum_{i>0}\left(JM_{i}M_{i+1}-\frac{1}{\beta}\log(1+2y\cosh(\beta(J(M_{i-1}+2M_{i}+M_{i+1})+h)))\right). \nonumber
\end{align}
On minimizing the free energy with respect to the magnetization we obtain mean field consistency equations for the  boundary and bulk magnetization. 
\begin{align}
&M_{0}=\frac{2y_{b}\sinh(\beta(2J_{b}M_{0}+JM_{1}+h_{b}))}{1+2y_{b}\cosh(\beta(2J_{b}M_{0}+JM_{1}+h_{b}))} \\\nonumber \\
&M_{i}=\frac{2y\sinh(\beta(J(M_{i-1}+2M_{i}+M_{i+1})+h))}{1+2y\cosh(\beta(J(M_{i-1}+2M_{i}+M_{i+1})+h))} \; (i>0)
\end{align}
For the bulk equations we impose the tricritical bulk parameters $h=0,\;y=1/4$ and \newline $\beta J=3/4$ as we are looking to model the tricritical Ising conformal field theory in the bulk. Then we invert them and on going to a continuum description $M_{i}=M(ia)$
we obtain
\begin{align}
a^{2}\partial^{2}M= \frac{4}{3}\log\left(\frac{2M}{1-M} + \sqrt{\frac{1 + 3M^{2}}{M^{2}-1}}\right)-4M.
\end{align}
This is equivalent to the bulk differential equation (\ref{nctribulkde}) obtained in the last section tuned to the tricritical point and with $\nabla^{2}=\partial^{2}$ as we assume no profile variation in the vertical direction.  The boundary equation on inversion and in the continuum picture becomes 
\begin{align}  
0=\mathrm{F}(M(0),y_{b})-2\beta J_{b}M(0)-\beta JM(a)-\beta h_{b}.\label{trimofabdyeq}
\end{align}
As with the Ising case we need to convert  the above boundary equation into an expression containing only the boundary magnetization that we relabel $M(0)=m$.  Thus we need to express the $M(a)$ term as a function of $m$, $M(a)=B(m)$.  We shall define this function in two different ways in exactly the same manner as was with the Ising model, the first as a series expansion  $B_{e}(m)$ and secondly as a numerical function $B_{n}(m)$.  We shall use the series expansion to analyze the nature of any critical points but as the expansion does not converge for $|m|<0.55$ we shall use the numerical function in order to produce any plots of $m(h_{b})$ or mean field superpotential $W_{mf}(m)$ in order that we can plot them across the whole range of $m$.  Inserting the expansion form $B_{e}(m)$ back into the equation (\ref{trimofabdyeq})
\begin{align}
\mathrm{F}(m,y_{b})-2\beta J_{b}m-\frac{3}{4}\cdot B_{e}(m)-\beta h_{b}=0 \label{tribdymaster}
\end{align}
we obtain a master boundary equation relating the boundary magnetization $M(0)=m$ and field $h_{b}$ that we can use to deduce the existence and nature of any critical points.  Though first we shall investigate the parameter limits we expect to correspond to the stable conformal boundary conditions ie limits where there is no fluctuation in the  boundary spins' occupancy.
\subsubsection{Zero boundary fluctuation limits}
In order to investigate the behaviour at the boundary parameter limits that one would expect to correspond to the stable conformal boundary states $(0)$ and $(\pm)$ we need the mean field occupancies $p_{0},p_{+}$ and $p_{-}$ that tell us the probability that the boundary spin  will occupy the $0$ or $\pm1$ levels.  We report them as 
\begin{align}
p_{+}=\frac{e^{\beta(K_{0}-(\Delta_{b}-h_{b}))}}{1+e^{\beta(K_{0}-(\Delta_{b}-h_{b}))}+e^{\beta(-K_{0}-(\Delta_{b}+h_{b}))}} \\ \nonumber \\
p_{-}=\frac{e^{\beta(-K_{0}-(\Delta_{b}+h_{b}))}}{1+e^{\beta(K_{0}-(\Delta_{b}-h_{b}))}+e^{\beta(-K_{0}-(\Delta_{b}+h_{b}))}}\\ \nonumber \\
p_{0}=\frac{1}{1+e^{\beta(K_{0}-(\Delta_{b}-h_{b}))}+e^{\beta(-K_{0}-(\Delta_{b}+h_{b}))}}
\end{align}
where $K_{0}=2 J_{b}M_{0}+JM_{1}$ and this term is always finite and thus will be dominated by infinite limits of $\Delta_{b}\pm h_{b}$  The conformal boundary RG flows end in infinite limits of the boundary parameters $h_{b}$ and $\Delta_{b}$ and thus we look for the mean field behaviour at the end of curves $h_{b}(\Delta_{b})$ in the $h_{b}\;\mathrm{vs}\;\Delta_{b}$ plane.  There are two cases to consider,  the first is when $\Delta_{b}<0$ and  when $\Delta_{b}\rightarrow -\infty$ we get $p_{0}\rightarrow 0$ for any curve $h_{b}(\Delta_{b})$.  Furthermore if $\mathrm{lim}_{\Delta_{b}\rightarrow-\infty}h_{b}(\Delta_{b})=\pm \infty$ we get $p_{\pm}\rightarrow1$ regardless of the how the curve $h_{b}(\Delta_{b})$ achieves this limit.  In order that we have a phase transition between the $\pm1$ boundary states we need neither $p_{\pm}\rightarrow 1$; the only curves that make this possible are ones such that $\mathrm{lim}_{\Delta_{b}\rightarrow-\infty}h_{b}(\Delta_{b})=\mathrm{constant}$, ie curves that are asymptotically parallel to the line $h_{b}=0$.  In this respect the line $h_{b}=0$ for $\Delta_{b}<0$ can be seen as a boundary between the $p_{\pm}=1$ stable limits.

Now we consider the case such that $\Delta_{b}\geq 0$.  When $\Delta_{b}\rightarrow\infty$, studying the occupation probabilities we see that for curves $h_{b}(\Delta_{b})$ st $\mathrm{lim}_{\Delta_{b}\rightarrow\infty}(\Delta_{b}- h_{b}(\Delta_{b}))=-\infty$ ie for curves whose asymptotic gradients are greater than $1$ we end up with $p_{+}\rightarrow1$. Similarly for curves st $\mathrm{lim}_{\Delta_{b}\rightarrow\infty}(\Delta_{b}+ h_{b}(\Delta_{b}))=-\infty$ ie ones whose asymptotic gradient is less than $1$ we get $p_{-}\rightarrow 1$ .For curves $\mathrm{lim}_{\Delta_{b}\rightarrow\infty}(\Delta_{b}\mp h_{b}(\Delta_{b}))=\infty$ ie 
ones whose asymptotic gradients are between and not including plus or minus one we get  $p_{0}\rightarrow1$.  From this we can draw the conclusion that any phase transition between the $0$ level and the $1$ level must occur at the endpoint of a curve that is parallel to the line $h_{b}=\Delta_{b}$.  Similarly for a phase transition between the $0$ and $-1$ levels must occur at the end of a curve asymptotically parallel to the line $h_{b}=-\Delta_{b}$.  Our next figure illustrates how the boundary parameter plane is divided into the respective regions.
\begin{figure}[H]
	\centering
		\includegraphics{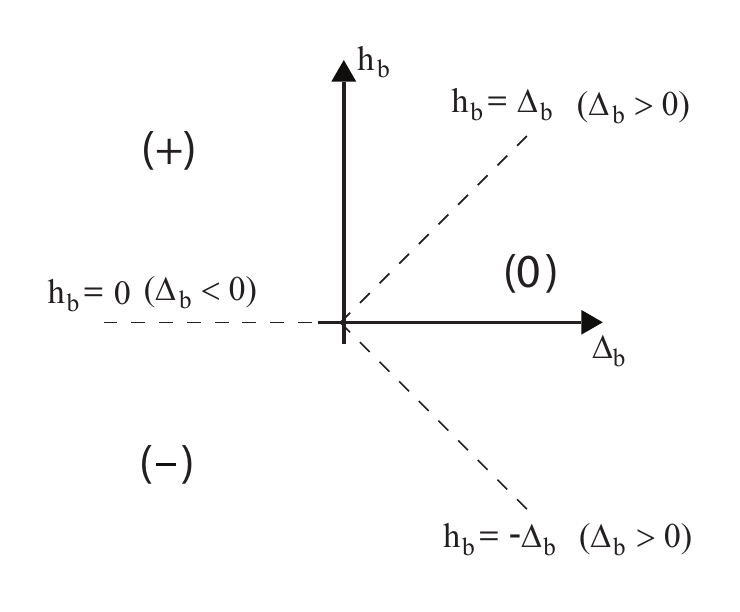}
	\label{fig:zerofluctuationregions}
	\caption{Regions of no boundary fluctuations}
\end{figure}
These results can be  taken to be exact because mean field  theory becomes exact in the regions of the  phase diagram far from any critical points.  As in the Ising case if we go back to our original lattice action (\ref{tribdyaction}) and take these stable limits, the terms $-h_{b}\sigma_{i,j}$ and $\Delta_{b}\sigma_{i,j}^{2}$ dominate completely the nearest neighbour terms.
\subsubsection{Boundary phase transitions}
\begin{figure}[H]
    \centering
   
    (a)
    {
        \includegraphics{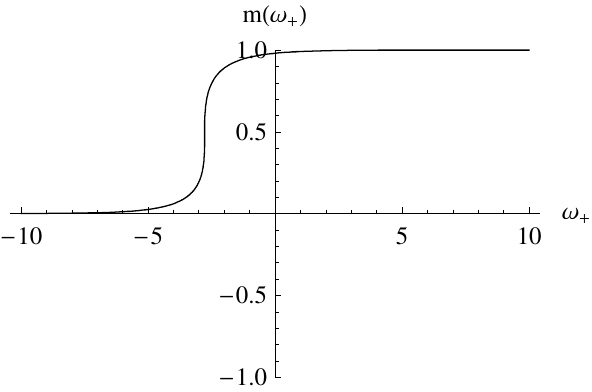}
        \label{fig:second_sub}
     } \;\;\;
   (b)
    {
        \includegraphics{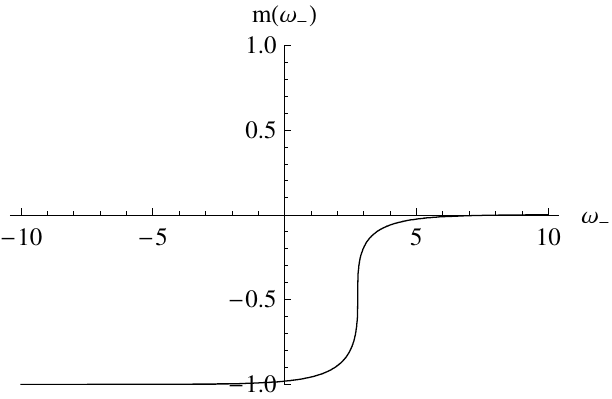}
        \label{fig:third_sub}
      }
    \caption{Plots of (a)$m(\omega_{+})$ at $(0+)$ limit (b) $m(\omega_{-})$ at $(0-)$ limit}
    \label{fig:trimixmohomega}
\end{figure}

In correspondence with the $(0\pm)$ Cardy boundary states we anticipate critical points in the limits of $(h_{b}\rightarrow \pm \infty, \Delta_{b} \rightarrow \infty)$ along a curve parallel to the line $h_{b}=\pm\Delta_{b}$.  We shall denote these lines $h_{b}=\pm\Delta_{b}+\omega_{\pm}$ (note that here we revert to using $\Delta_{b}$ instead of $y_{b}=e^{-\beta \Delta_{b}}$).  Taking these respective limits $(h_{b}\rightarrow \pm \infty, \Delta_{b} \rightarrow \infty)$ our master boundary equation takes the form 
\begin{align}
2\mathrm{arctanh}(2m\mp1)-2\beta J_{b}m -\frac{3}{4}B_{e}(m)-\beta \omega_{\pm}=0.
\end{align}
As in the bulk we are unable to expand this equation around $m=0$  as the arctanh term diverges there, in the bulk we obtained an exact value for the critical value of $\bar{m}=1/2$ and we could then retrospectivley expand around that.  In the boundary case as we shall see at best we are able only able to obtain an approximate numerical value for the critical value of the boundary magnetization and as a result we cannot expand around it. Thus we use the point of inflection condition of $\partial_{m}h_{b}=\partial^{2}_{m}h_{b}=0$ on the full mean field equation as the condition for criticality.  Obviously we cannot consider all the terms in the expansion $B_{e}(m)$ and so we truncate it to the first twenty terms as including more makes no difference to the numerical answers obtained.  In doing so we obtain critical points at these limits with the parameters taking the critical values $m=\pm0.477,\beta J_{b}=1.76,\omega_{\pm}=\mp2.77\;\mathrm{(3sf)}$.  In figure (\ref{fig:trimixmohomega}) we plot the behaviour of $m(\omega_{\pm})$ and $W(m)$ at the respective $(0\pm)$ limits just explored.
These plots indicate the fact that $\omega_{\pm}$ are the parameters defining the boundary conditions once we have taken the limits $\Delta_{b}\rightarrow \infty,\;h_{b}/\Delta_{b}\rightarrow\pm1$ that correspond to the respective $(0\pm)$ conformal boundary states.

\begin{figure}[H]
    \centering
   
    (a)
    {
        \includegraphics{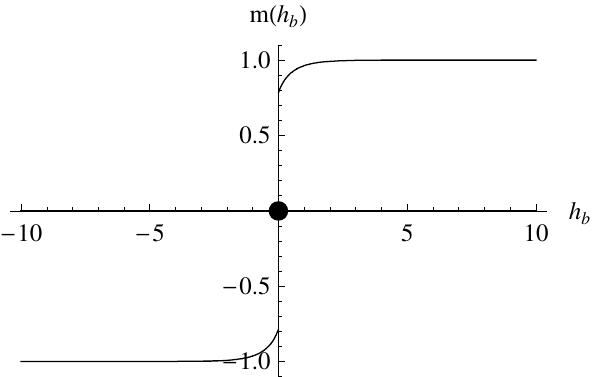}
        \label{fig:second_sub}
     } \;\;\;
    (b)
    {
        \includegraphics{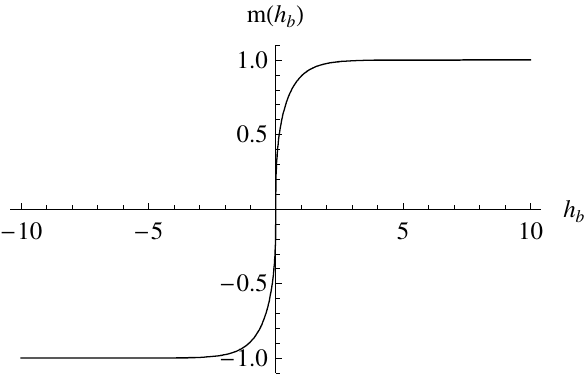}
        \label{fig:third_sub}
      }
    \caption{Plots of $m(h_{b})$ when $\Delta_{b}\rightarrow \infty$ (a) for $\beta J_{b}=2/5$, the black dot indicates the zero solution we take as being physical in our interpretation (b)for $\beta J_{b}=1/8$}
    \label{fig:trisupmofh}
\end{figure}

The conformal space of flows would suggest that a first order transition should be present in the limit $(\Delta_{b} \rightarrow - \infty,h_{b}=0)$.  From our investigation into the stable limits we concluded that there would  be a phase transition from the $\{+\}$ level to the $\{-\}$ level in this limit of $(\Delta_{b} \rightarrow - \infty,h_{b}=0)$. In this limit our master boundary equation (\ref{tribdymaster}) becomes 
\begin{align}
\mathrm{arctanh}(m)-2\beta J_{b}m -\frac{3}{4}B_{e}(m)-\beta h_{b}=0
\end{align}
this can be expanded for small $m$ resulting in the Landau expansion
\begin{align}
0=-\beta h+(\frac{1}{4}-2\beta J_{b})m+\left(\frac{1}{3}+\frac{3}{4}\sqrt{\frac{3}{5}}\right)m^{3}+O(m^{5}).
\end{align}
Here one obtains a critical phase transition when $h_{b}=0$ and $\beta J_{b}=1/8$. However as previously stated results from conformal field theory tell us that there should be a first order transition (discontinuous jump in magnetization) present here. In order for a first order transition to occur in our mean field theory treatment we would require $\beta J_{b}>1/8$ and one gets three solutions to the mean field equation, one being $m=0$ and the others $m=\pm \alpha$ where $0<\alpha<1$.  Normally one would not consider the zero solution as physical as it sits atop a maximum of the potential, however spontaneous  magnetization is not possible at a one dimensional boundary and thus in our modeling we will take $m=0$ to be the physical solution for $h_{b}=0$ and when a small $h_{b}$ is switched on the boundary magnetization jumps to either $m=\pm \alpha$ depending on the sign of the  perturbation. In this sense we effectively model the exact picture coming from conformal field theory.  It should  be noted however that in our mean field picture if one wanted a maximal first order transition ie a finite jump to either $m=\pm1$ then $\beta J_{b}$ would have to tend to infinity or conversely the temperature would have to be zero. In figure (\ref{fig:trisupmofh}) we produce full plots of both  $m(h_{b})$ and $W(m)$ setting $\beta J_{b}=2/5$, quite an arbitrary value but one that illustrates the possibility of a first order transition.

\begin{figure}[H]
    \centering
   
    (a)
    {
        \includegraphics{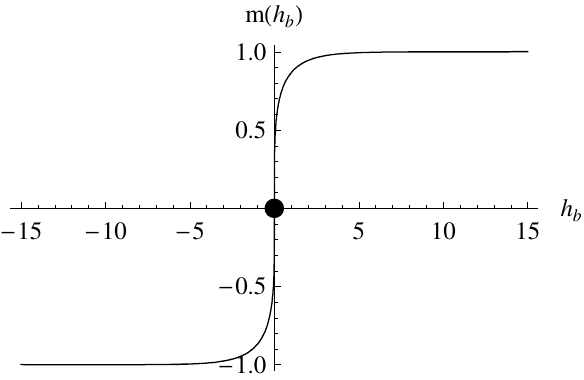}
        \label{fig:second_sub}
     } 
    (b)
    {
        \includegraphics{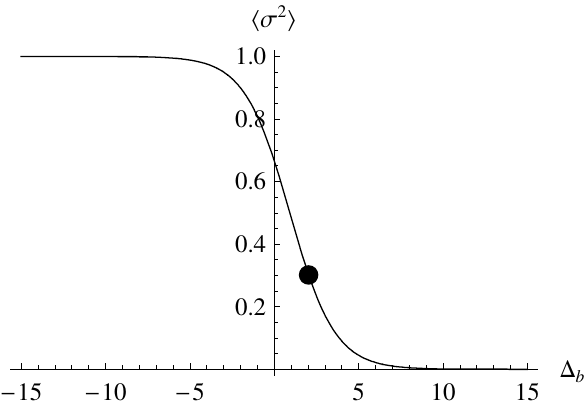}
        \label{fig:third_sub}
      }
    \caption{(a)$m(h_{b})$ through tricritical point ($h_{b}=0,\Delta_{b}=-4/3\log(y_{bcrit})=2.04\mathrm{(3sf)}$) indicated by a black spot   (b) $\left\langle \sigma^{2}\right\rangle(\Delta_{b})$ through tricritcal point again indicated by a black spot}
    \label{fig:tritri}
\end{figure}

\begin{figure}[H]
	\centering
\includegraphics{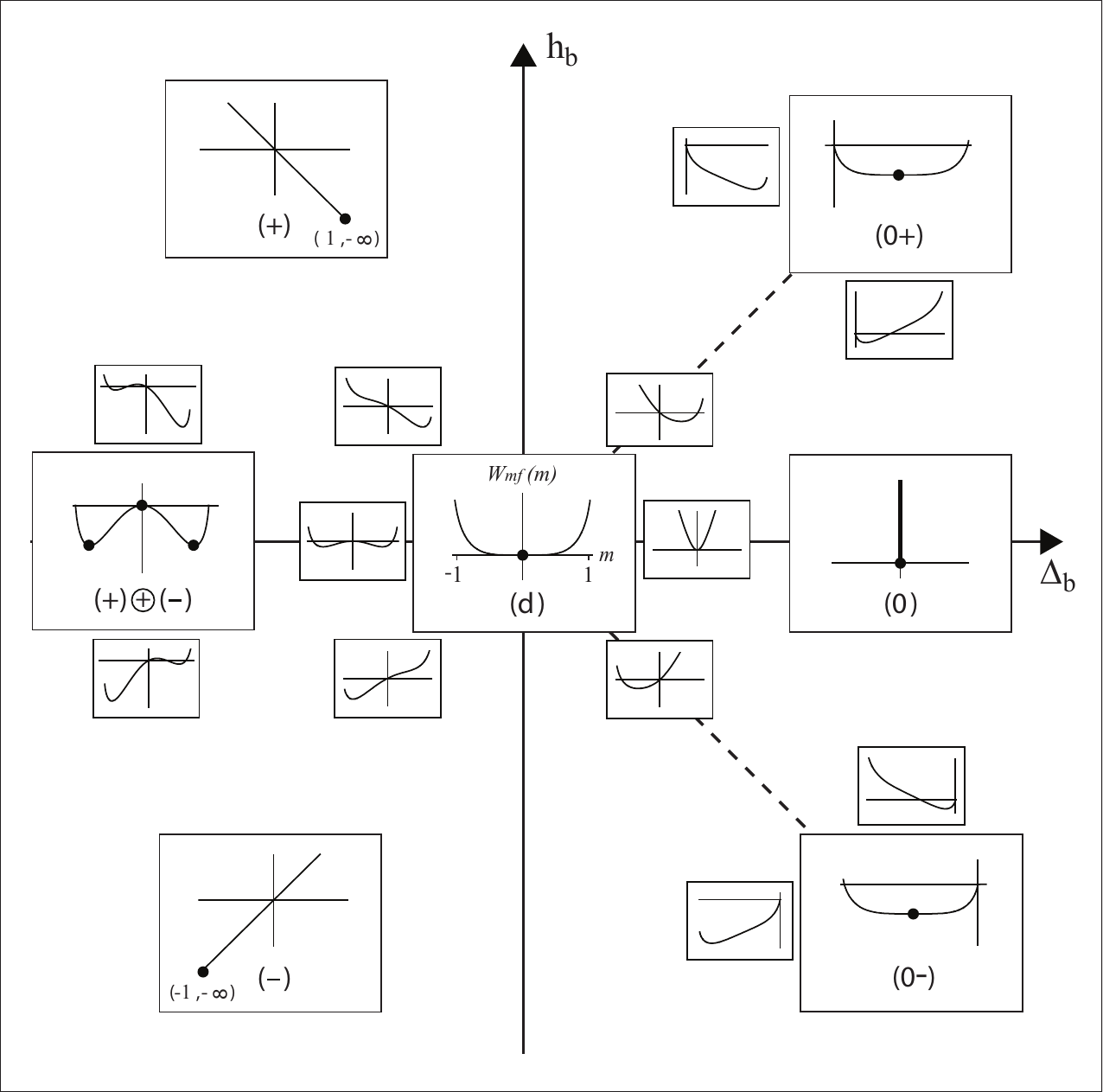}
  \caption{Boundary tricritical Ising model phase diagram}
	\label{fig:overalltri1}
\end{figure}

We have considered the phase transitions occurring at the limits of lines parallel to $h_{b}=\pm\Delta_{b},0$ and at most one could obtain a second order critical point along said lines.  We expect a phase transition at roughly the intersection of these three lines in analogy with the bulk tricritical point.  Expanding the lhs of our master boundary equation (\ref{tribdymaster}) for small $m$ we obtain
\begin{align}
0=&-\beta h_{b}+ (\frac{1}{2y_{b}}-2\beta J_{b} +\frac{1}{4})m+\left(\frac{(4y_{b}-1)(1+2y_{b})^{2}}{48y_{b}^{3}}+\frac{3}{4}\sqrt{\frac{3}{5}}\right)m^{3} \label{trilan}\\ &+\left(\frac{(3-18y_{b}+32y_{b}^{2})(1+2y_{b})^3}{1280y_{b}^{5}}-\frac{9}{560}(56+5\sqrt{15})\right)m^{5}+\mathrm{O}(m^{7})\nonumber
\end{align}
this is critical if $\beta J_{b}=(2+y_{b})/8y_{b}$ and $h_{b}=0$ and tricritical if
\begin{align}
y_{b}=y_{btri}=&\frac{1}{240+108\sqrt{15}}\left(\left(1528200+388800\sqrt{15}-16200\sqrt{17361+4528\sqrt{15}}\right)^{\frac{1}{3}}\right.\nonumber\\
&+\left.6\cdot5^{\frac{2}{3}}\left(283+72\sqrt{15}+3\sqrt{17631+4528\sqrt{15}}\right)^{\frac{1}{3}}-60\right)\nonumber \\ \nonumber \\
=&0.216\;\mathrm{(3sf)}
\end{align}
This tricritcal phase transition we identify with the conformal $(d)$ state.  We now illustrate the tricritcal phase transition in figure (\ref{fig:tritri})
We have included a plot of $\left\langle \sigma^{2}\right\rangle$ versus $\Delta_{b}$ for $h_{b}=0$.  This we hope illustrates the qualititive nature of the conformal boundary flows from the $(d)$ state to either of the $(+)\oplus(-)$ and $(0)$ states.  With $\left\langle \sigma^{2}\right\rangle=1$ corresponding to the state $(+)\oplus(-)$ and $\left\langle \sigma^{2}\right\rangle=0$ corresponding to the $(0)$ state.  In figure (\ref{fig:overalltri1}) we summarize the boundary phase diagram using our full mean field superpotential $W_{mf}(m)$ obtained by integrating the numerical form of our boundary equation as in the Ising case.
In the center we have the tricritical point we identify with the conformal boundary $(d)$ state, which similarly to the  bulk occurs at the intersection of three lines of critical points.  We only explicitly worked out the case for the infinite limits of said critical lines but on investigation one finds that there are critical points along the whole of the lines.  The limit of the two of these critical lines with $\Delta_{b}\rightarrow \infty$ we identify with the $(\pm0)$ conformal states.  The $\Delta_{b}\rightarrow -\infty$ critical line we plot with $\beta J_{b}>1/8$ in order to give a first order transition as is expected from conformal field theory and we identify this limit with the $(+)\oplus(-)$ conformal boundary state.    Any other curves ending in infinite limits of the boundary parameters we have shown will end up at stable boundary conditions that we identify with the $(0),(\pm)$ conformal boundary states.  The smaller plots in the figure relate to perturbing the boundary parameters slightly about the phase transition in question.
\section{Comparison with work of Cappelli et al}
In the case of the Ising model considered in section 2 we saw that Cappelli's superpotential agreed with our Landau approximation to the boundary mean field equations if we tuned $\beta J_{b}$ to it's critical value.  However such an approximation is only valid for small  values of the  order parameter $m$, in the regions  $h_{b}\rightarrow \pm \infty$ we get $m \rightarrow \pm 1$ of the phase diagram where such a Landau approximation of the full mean field equations would not hold and furthermore in these regions mean field equations become exact. If one does take the Landau approximation to hold in all regions one would get a picture where $m\rightarrow \pm \infty$ and it would be  these limits that Cappelli identifies with the conformal boundary states $(\pm)$ arguing that in the conformal picture any nonzero boundary magnetization would get renormalized to infinity and thus it is natural to identify these limits.

In the case of the boundary tricritical Ising model Cappelli put forward the following polynomial potential $W_{cap}(m)$
\begin{align}
W_{cap}(m)=bm-a\frac{m^2}{2}+\frac{m^{4}}{4}=-h_{b}m+\delta_{b}m^{2}+\frac{m^{4}}{4}
\end{align}
where in the second equality we relabel his parameters $b=-h_{b}$ and $\delta_{b}=-a/2$ for direct comparison with our work.  He identifies certain points and limits in the boundary parameter space $h_{b}vs\delta_{b}$ with the the Cardy boundary states as we have done.  In his analysis he states that the boundary value $m$ should be given as the stationary points of his superpotential, the stationary equation $\partial_{m}W_{cap}=0$ is the following cubic
\begin{align}
0=-h_{b}+2\delta_{b}m+m^{3}
\end{align}
one can differentiate again to find the equation that defines a point of inflection for $W_{cap}$, $\partial_{m}^{2}W_{cap}=0$, this results in $m=\pm(-2\delta_{b}/3)^{1/2}$ at these points of inflection and on substitution of this value of $m$ back into the stationary equation we obtain the relation between $h_{b}$ and $\delta_{b}$ as is necessary for an inflection point.   One obtains $h_{b}=\pm2(-2\delta_{b}/3)^{3/2}$, for $\delta_{b}<0$ these wings define boundaries in the parameter space, to the right of the wings there is only one real solution and two complex conjugate solutions to the cubic stationary equation and the real solution will be a minimum. On the wings and to the left of them there will be three real solutions with different degrees of degeneracy.  To the left of the wings there are three distinct real solutions, on the line $h_{b}=0, \delta_{b}<0$ one of the solutions will be $m=0$ and the other two will be symmetric about this point $m=\pm \alpha$. Please refer to the next figure for the following discussion, (the wings are dotted lines).
\begin{figure}[H]
	\centering
\includegraphics{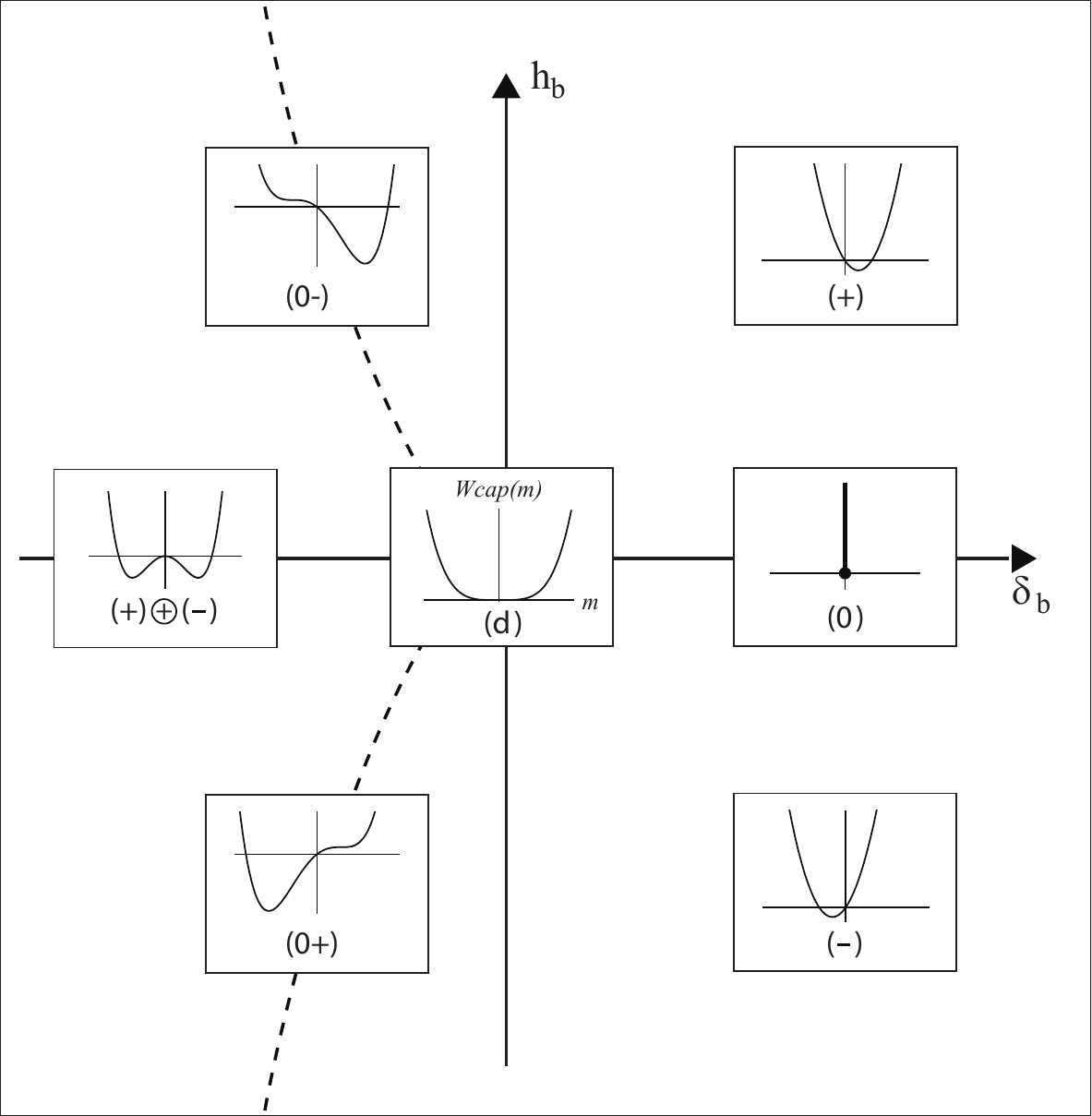}
	\label{fig:overall}
	\caption{Tricritical Ising model phase diagram of Cappelli et al}
\end{figure}
On the wings themselves (not including origin $h_{b}=\delta_{b}=0$) two of the distinct solutions present just to the left of the wing will have coincided on going onto the wing and the coincident solution represents the point of inflection. The third real solution on the wing will be a minimum.  At the origin $h_{b}=\delta_{b}=0$ there will be three coincident solutions of $m=0$.  Cappelli uses the degeneracy of the stationary points as the guiding light in his analysis as to what points or limits to identify with which Cardy states.  The reason being is that there can be a nice match between the number of relevant fields at the given Cardy states and $(n-1)$ where $n$ is the degeneracy of the points/limits Cappelli chooses.  Accordingly as at the origin $n-1=2$ and as there are $2$ relevant fields at the $(d)$ boundary state he makes the  identification of the origin $h_{b}=\delta_{b}=0$ with the free boundary state.  As  $n-1=1$ along the wings and there is one relevant field at the $(\pm0)$ states he identifies them with the limits $(h_{b}\rightarrow\pm\infty,\delta_{b}\rightarrow -\infty)$ along the wings.  To the right of the wings $n-1=0$ and he thus identifies certain infinite limits of $h_{b}$ and $\delta_{b}$ here with the the stable boundary states $(\pm),(0)$.  He does not however specify along which curves one should take these limits.  Only that any limit $(h_{b}\rightarrow\pm\infty,\delta_{b}\rightarrow\infty)$ should be identified with the respective $(\pm)$ states and that the limit $(h_{b}=0,\delta_{b}\rightarrow\infty)$ should be identified with the $(0)$ state.  For the Cardy state superposition $(+)\oplus(-)$ Cappelli identifies the limit  $(h_{b}=0,\delta_{b}\rightarrow-\infty)$ however the match between $n-1=2$ here and the number of relevant boundary fields does not hold, possibly one could say identify solutions of the form $m=\pm\alpha$.

We think there are a number of issues with this picture of the boundary phase diagram and we will now compare Cappelli's picture with our full mean field picture and argue that ours gives the correct physical picture.  Starting with the  phase transition corresponding to the $(d)$ state in our picture we obtain the polynomial expansion $W_{Lan}(m)$ by integrating the our Landau expanded boundary equation (\ref{trilan})
\begin{align}
W_{Lan}=&-\beta h_{b}m+ \frac{1}{2}(\frac{1}{2y_{b}}-2\beta J_{b} +\frac{1}{4})m^{2}+\frac{1}{4}(\frac{(4y_{b}-1)(1+2y_{b})^{2}}{48y_{b}^{3}}+\frac{3}{4}\sqrt{\frac{3}{5}})m^{4}\\ &+\frac{1}{6}\left(\frac{(3-18y_{b}+32y_{b}^{2})(1+2y_{b})^3}{1280y_{b}^{5}}-\frac{27}{2240}(56+5\sqrt{15})\right)m^{6}\nonumber
\end{align}
the coefficients in this expansion have a complicated functional dependence on $\Delta_{b}$, for a more direct comparison we can expand the coefficients around the tricritical values of the parameters
\begin{align}
\beta J_{b}=\frac{y_{btri}+2}{8y_{btri}}+\frac{\epsilon}{2}\;\;\;\;\beta\Delta_{b}=-\log(y_{btri})+\frac{\delta}{2}
\end{align}
to result in
\begin{align}
W_{Lan}=-h_{b}m+(\delta_{b}-\epsilon)m^{2}-0.474\delta_{b} m^4+(0.293+1.04\delta_{b})m^{6}
\end{align}
the decimals in the coefficients can be given as very long exact answers but we give present them as decimals given to (3sf).  Now our term $\delta_{b}$ is directly comparable to Cappelli's and we see that if we tune $\epsilon=0$ then the coefficients of our $m^{2}$ terms  become the same, but again there is further coupling at the higher order terms.  This entails that we obtain a tricritical point in analogy to  the bulk picture when we tune $\delta_{b}=0$ whereas Cappelli's analysis predicts a critical point.  It is important to state that by construction our potential $W_{Lan}$ is only valid for values of the boundary parameters close to the tricritical point.

The next issue is with the $(\pm0)$ limits, in Cappelli's picture he has the unphysical behaviour in that he has the $(0+)$ state occurring for a negative $h_{b}$, whereas the $(+)$ state occurs for positive $h_{b}$, meaning that the magnetization field changes the nature of its alignment with the external field at different regions in the phase diagram.  In our picture we obtain consistent alignment between the magnetization and external field throughout the phase diagram. Furthermore we found that we could n't even obtain a polynomial Landau expansion and had to rely on the full mean field equations at these phase transitions, even in the bulk case where we could obtain a polynomial expansion we had to take the limits $(h_{b}\rightarrow\pm\infty,\Delta_{b}\rightarrow\infty)$ first in the full mean field equations and then make the expansion.  There was no way we could obtain a polynomial expansion that held in all regions of the phase diagram.  This is to be expected as the crux of the Landau approximation is to expand around a small value of the order parameter and in the phase diagram of the bulk or boundary models there are regions where the order parameter is not small.  The final issue with Cappelli's picture is that our analysis showed exactly that any phase transition $(\pm0)$ that occurred at the infinite endpoints of curves in the $h_{b}\mathrm{vs}\Delta_{b}$ plane had to occur at the endpoints of lines asymptotically parallel to the lines $h_{b}=\pm\Delta_{b}$.  Cappelli says the $(\pm0)$ occur at the endpoints of the wings $h_{b}=\pm2(-2\delta_{b}/3)^{3/2}$, unless there is some very particular coupling between the vacancy density as defined in the lattice Hamiltonian and the quadratic coefficient $\delta_{b}$ in Cappelli's picture it is hard to see how the wings would n't end in stable limits.   Unfortunately Cappelli makes no attempt to link the parameters in his superpotential with the tricritical Ising Hamiltonian.

A final issue with Cappelli's analysis is that he does not state along which curves one would obtain the $(\pm)$ stable boundary states, he only says that they should be associated with the asymptotic regions $(h_{b}=\pm\infty,\delta_{b}=\infty)$.  The trouble is with his superpotential one would need to actually specify along which curves one should reach these asymptotic regions in. For instance if one reaches these limits along curves of the form $h_{b}= \pm\delta_{b}^{\alpha}$ with $0<\alpha<1$ then the stationary equation becomes 
\begin{align}
\mp\delta_{b}^{\alpha}+2\delta_{b}m+m^{3}=0
\end{align}
dividing through by $\delta_{b}$ one gets 
\begin{align}
\mp\delta_{b}^{\alpha-1}+2m+\frac{m^{3}}{\delta_{b}}=0
\end{align}
in the limit $\delta_{b}\rightarrow\infty$ and with $0<\alpha<1$ the unique solution is $m=0$ which is the value associated to the $(0)$ and not the $(\pm)$ states. 
\\
\section{Conclusion}
In conclusion by extending the existing techniques of mean field theory we have presented a qualititive yet comprehensive picture of the boundary phase diagram of both Ising and tricritical Ising lattice models.  Both showing agreement with the known pictures from conformal field theory.  Furthermore we have found disagreement with the work of Cappelli et al. The crux of which is due to our equations are derived from first principles from the underylying lattice Hamiltonians whereas their work is based on what could apparentley be only an coincidental agreement between the nested pattern of RG flows in the boundary minimal models and the nested nature of the singular points in Arnold's theory of singularities.  In taking this route, Cappelli et al. we think miss the true nature of the boundary critical points and go against the original principle of Landau theory, which is to only expand about small values of the order parameter.  Finally the full mean field equations give useful information on the level probabilities throughout the phase diagram, which we found useful in defining the stable $(0)$ and $(\pm)$ regions and in these regions mean field theory becomes exact.

One could extend our work to all higher critical Ising models that would model the behaviour of their corresponding minimal models to give a qualititive agreement with the work of K.Graham \cite{graham}.  Further as mentioned in the introduction that the principle motivation for the development of these techniques is to extend them to the case of conformal defects in minimal models as investigated in \cite{Watts1} and \cite{Watts2} and to find nature of new defects.
\section{Acknowledgments}
I would like to thank G.Watts for invaluable discussions and comments.  I have been supported by STFC studentship number PPA/S/S/2005/04103, and partially supported by STFC grant ST/G000395/1.

\end{document}